\documentclass[9pt,twocolumn,twoside]{osajnl}
\journal{josab} % Choose journal (ao, aop, josaa, josab, ol, pr)

\setboolean{shortarticle}{false}
\usepackage{siunitx} % for SI units
\usepackage{graphicx}
\usepackage{comment} % comment block
\usepackage{multirow}
\usepackage{chemformula} % Formula subscripts using \ch{}
\usepackage[T1]{fontenc} % Use modern font encodings
\usepackage{dcolumn}% Align table columns on decimal point
\usepackage{bm}% bold math
\usepackage{siunitx}
\usepackage{ulem}
\usepackage{natbib}
\usepackage{graphicx}
\usepackage{dcolumn}
\usepackage[english]{babel}
\usepackage{amsmath}
\usepackage{amssymb}
\usepackage{tabularx}
\usepackage{multirow}
\usepackage[all]{xy}
\usepackage{appendix}
\usepackage{xspace}
\usepackage{color}

\newcommand{\ket}[1]{\ensuremath{|#1\rangle}\xspace}

\title{An efficient multimode vectorial nonlinear propagation solver beyond the weak guidance approximation}

\author[1,*]{Pierre B\'ejot}

\affil[1]{Laboratoire Interdisciplinaire Carnot de Bourgogne (ICB), UMR6303 CNRS--Universit\'e de Bourgogne, 21078 Dijon, France}

\affil[*]{Corresponding author: pierre.bejot@u-bourgogne.fr}

\begin{abstract}
In this article, we present an efficient numerical model able to solve the vectorial nonlinear pulse propagation equation in circularly symmetric multimode waveguides. The algorithm takes advantage of the conservation of total angular momentum of light upon propagation and takes into account the vectorial nature of the propagating modes, making it particularly relevant for studies in ring-core fibers. While conventional propagation solvers exhibit a computational complexity scaling as $N_{\textrm{mode}}^{4}$ where $N_{\textrm{mode}}$ is the number of considered modes, the present solver scales as $N_{\textrm{mode}}^{3/2}$. As a first example, it is shown that orbital angular momentum modulation instability processes take place in ring-core fibers in realistic conditions. Finally, it is predicted that the modulation instability process is followed by the appearance of breather-like angular structures.
\end{abstract}

\setboolean{displaycopyright}{true}

\begin{document}

\maketitle

\section{Introduction}
\label{sec:intro}
Advancements in optical fiber technology have sparked renewed interest in the potential use of multimode waveguides to address limitations inherently associated with monomode propagation. The abundance of potential waveguide architectures now available opens doors to exploring an extremely rich variety of propagation dynamics \cite{KrupaReview,WrightNatPhys}. Solitons \cite{CrosignaniSoliton,HasegawaSoliton,CrosignaniSoliton2,YuSoliton,RaghavanSoliton,RenningerSoliton} and associated resonant radiations through geometric parametric instabilities \cite{KrupaResonantRadiations}, invariant space-time wavepackets \cite{BejotPRE,BejotPRL,BejotHelicon,BejotQuadrics,StefanskaACS}, self-cleaning \cite{LiuCleaning,KrupaCleaning} and related condensation/thermalization \cite{Aschieri,PicozziPhysReports,WuThermo,Pourbeyram,WuThermalizationOAM,PodivilovOAMThermalization} phenomena are some nonlinear fascinating effects, among others, taking place in multimode fibers that have been intensively studied during the last few years. Beside these nonlinear phenomena, the use of multimode fibers for increasing the number of data-transmitting channels lets envision the possibility to overcome the expected data-carrying capacity crunch of single-mode fibers. In this regard, specially designed multimode optical fibers, called ring-core fibers, have been developed so as to minimize spatial-mode couplings \cite{BozinovicScience,Brunet,RamachandranRingCore}. The originality of these fibers lies in the modal content they support. First, contrary to conventional multimode fibers, these fibers only support one single kind of ring-shaped radial modes (p-modes) with well separated propagation constants, which has the advantage to strongly limit mode coupling. The other strong advantage of the latter property is that this reduces the initially full 3D+1 problem ([$r,\theta,t,z]$, where ($r,\theta,z$) are the cylindrical coordinates, $z$ being the direction along which the light propagates, and $t$ stands for time) into a far simpler 2D+1 ($\theta,t,z$) problem. In a theoretical point of view, this dimensionality reduction lets envision the possibility to study the existence of new kind of solutions, otherwise impossible to find in more complex waveguides. Another important property of the modes supported by these fibers is that they are also eigenvectors of the total angular momentum (TAM) operator. This is in fact true not only for ring-core fibers but also for any circularly symmetric waveguides. However, the peculiarity of the ring-core fiber modes lies in their polarization content. While modes of conventional fibers are in very good approximation circularly polarized (excepted for the special case of transverse-electric and transverse-magnetic modes), the ones of ring-core fibers are extremely \textit{vectorial} \cite{GreggMomentumConservation}, i.e., their polarization contents strongly vary within their transverse sections. Theoretically speaking, the immediate consequence of this point is that scalar propagation equation does not hold anymore in such a system. Accordingly, if one wants to simulate the (nonlinear) propagation within these fibers, one necessarily has to use vectorial propagation solvers, which obviously increases the complexity of the simulations. The latters can become strongly computationally demanding, in particular when the waveguide is massively multimode. In this context, it is then of paramount importance to develop fast and accurate full vectorial models able to capture as close as possible the nonlinear propagation dynamics in waveguiding structures potentially supporting hundreds of modes.\\
Recently, we developed a numerical algorithm able to efficiently solve the scalar version of the nonlinear unidirectional pulse propagation equation (UPPE) \cite{BejotPRE}. This algorithm, valid in the weak guidance approximation, was shown to be not only more accurate since based on UPPE, but also far faster \cite{TarnowskiJOSAB} than the classical algorithm solving the multimode generalized nonlinear Schrodinger equation (MM-GNLSE), even if the latter was massively parallelized \cite{WrightSimul}. This is because the complexity of MM-GNLSE scales approximately as $N^4$ \cite{Poletti2008}, where $N$ is the number of modes to handle while, as shown below, the developed algorithm approximately scales as $N^{3/2}$. Note however, that the superiority of our algorithm on MM-GNLSE ones in terms of calculation speed only holds when the number of modes becomes higher and higher. In this article, we present a vectorial version of our algorithm that remains valid beyond the weak guidance approximation, i.e., that takes into account the vectorial nature of the modes. We restrict this study to waveguides presenting a cylindrical symmetry around the propagation axis. In the first section, after recalling the propagation equation, we discuss the general properties of vectorial modes and their intimate link with the TAM of light. In particular, we will show how the conservation of TAM in circularly symmetric waveguide can be advantageously used for constructing a fast-modal transform, i.e., a transformation allowing to go back and forth between a modal representation of the field and its representation in the direct space. In the second section, we will present the developed algorithm for solving in an efficient way the vectorial nonlinear UPPE equation. Finally, in the last section we will present a numerical example. In particular, it is predicted, for the first time, that a moderately intense continuous field propagating in a ring-core fiber can evolve towards an angular breather-like structure.
\section{Linear propagation equation and mode structure of circularly symmetric waveguides}
\label{sec:Equation}
\subsection{Propagation equation}
We consider here an isotropic non-magnetic medium without current and charges, whose linear susceptibility $\chi^{(1)}$ is inhomogeneous in space. At this point, since this section is devoted to study the modal structure of waveguides, we consider a linear propagation regime.
The Maxwell's equations read:
\begin{eqnarray}
\begin{aligned}
&\overrightarrow{\nabla}\times\overrightarrow{E}=-\partial_t\overrightarrow{B},\, \overrightarrow{\nabla}\cdot\overrightarrow{D}=0&\\
&\overrightarrow{\nabla}\times\overrightarrow{B}=\mu_0\partial_t\overrightarrow{D},\,\overrightarrow{\nabla}\cdot\overrightarrow{B}=0,&
\end{aligned}
\end{eqnarray}
where $\mu_0$ is the vacuum permeability, $\overrightarrow{E}$ and $\overrightarrow{B}$ are the electric and magnetic fields respectively. In the frequency space, the electric displacement field $\overrightarrow{\tilde{D}}$ respects
\begin{equation}
\overrightarrow{\tilde{D}}=\epsilon_0n^2\overrightarrow{\tilde{E}},
\end{equation}
where $n=\sqrt{1+\chi^{(1)}}$ is the frequency- and spatially-dependent refractive index and $\epsilon_0$ is the vacuum permittivity.
Since $\overrightarrow{\tilde{D}}$ has a divergence equal to zero, it implies:
\begin{equation}
\overrightarrow{\nabla}\cdot\overrightarrow{\tilde{E}}=-\overrightarrow{\nabla}\left[\textrm{log}\left(n^2\right)\right]\cdot\overrightarrow{\tilde{E}}.
\label{Eq_Divergence}
\end{equation}
The electric field propagation equation then reads in the frequency space
\begin{equation}
\overrightarrow{\Box}\overrightarrow{\tilde{E}}=\overrightarrow{0},
\label{EqPropagationVectorielle}
\end{equation}
with
\begin{equation}
\overrightarrow{\Box}\overrightarrow{\tilde{E}}={\nabla}^2\overrightarrow{\tilde{E}}+\frac{n^2\omega^2}{c^2}\overrightarrow{\tilde{E}}+\overrightarrow{S}[\overrightarrow{\tilde{E}}],
\label{BoxOperator}
\end{equation}
where $\omega$ is the angular frequency, $c$ is the speed of light in vacuum, and
\begin{equation}
\overrightarrow{S}[\overrightarrow{\tilde{E}}]=\overrightarrow{\nabla}\left[\overrightarrow{\nabla}\left(\textrm{log}\,n^2\right)\cdot\overrightarrow{\tilde{E}}\right].
\end{equation}
For a circularly symmetric waveguide (supposed homogeneous all along $z$), the refractive index only depends on $r$ in the cylindrical coordinates system ($r$,$\theta$,$z$) and, for simplicity, we define $f(r)=\textrm{log}\,n^2$.
In this case, one has
\begin{equation}
\overrightarrow{\nabla}f(r)=\partial_rf(r)\overrightarrow{e}_r.
\end{equation}
so that
\begin{equation}
\overrightarrow{S}[\overrightarrow{\tilde{E}}]=\overrightarrow{\nabla}\left(\overrightarrow{\nabla}f\cdot\overrightarrow{\tilde{E}}\right)=
\begin{pmatrix}
\partial^2_rf\ \tilde{E}_r+\partial_rf\ \partial_r \tilde{E}_r\\
\frac{1}{r}\partial_rf\ \partial_\theta \tilde{E}_r\\
\partial_rf\ \partial_z\tilde{E}_r
\end{pmatrix},
\label{Eq1}
\end{equation}
where the electric field $\overrightarrow{\tilde{E}}$ is decomposed in cylindrical coordinates as
\begin{equation}
\overrightarrow{\tilde{E}}=\tilde{E}_r\overrightarrow{e}_r+\tilde{E}_\theta\overrightarrow{e}_\theta+\tilde{E}_z\overrightarrow{e}_z.
\end{equation}
Physically speaking, the operator $\overrightarrow{S}[\overrightarrow{\tilde{E}}]$ is directly related to the spin-orbit coupling phenomenon, as we will discuss a bit latter.
\subsection{Total angular momentum of light: eigenvectors and eigenvalues}
Let us consider the operators $L_z$ and $S_z$ acting on the electric field as:
\begin{equation}
L_z=-i\frac{\partial}{\partial\theta}I_d,\,S_z=i
\begin{pmatrix}
0 & -1\\
1 & 0
\end{pmatrix}
,
\end{equation}
where $I_d$ is the two-dimensional identity matrix, and $\theta$ the polar angle. The operator $L_z$ (resp. $S_z$) corresponds to the projection along $z$ of the orbital (resp. spin) angular momentum of light. They can be seen as infinitesimal generators of rotations around the $z$ axis of the field amplitude and polarization, respectively. While the eigenvectors of $S_z$ correspond to the left and right circularly polarized fields $\left(\ket{\sigma_+},\ket{\sigma_-}\right)$ with respective eigenvalues $\left(1,-1\right)$, the eigenvectors of $L_z$ are fields owing an helical wavefront $e^{i\ell\theta}$ ($\ell\in \mathbb{Z}$) with eigenvalues $\ell$.\\
If one now considers the TAM of light, any eigenvectors of $J_z=L_z+S_z$ write
\begin{equation}
\ket{\Phi_{j_\ell}}=e^{i\ell\theta}\left[a_+\ket{\sigma_+}+a_-e^{2i\theta}\ket{\sigma_-}\right],
\label{EigenVectorTAM}
\end{equation}
with corresponding eigenvalues $j_\ell=\ell+1$ ($\ell \in \mathbb{Z}$) and $(a_+,a_-)$ the respective complex amplitudes of the two circularly polarized components. In particular, for a given $j_\ell$, one can define the two orthonormal eigenvectors $\ket{\Psi_{\pm,j_\ell}}$
\begin{equation}
\ket{\Psi_{\pm,j_\ell}}=\frac{e^{i\ell\theta}}{\sqrt{2}}\left[\ket{\sigma_+}\pm e^{2i\theta}\ket{\sigma_-}\right],
\label{basisSet}
\end{equation}
that form a bi-dimensional basis of representation of any eigenvectors of $J_z$ with corresponding eigenvalue $j_\ell=\ell+1$. In particular, one has:
\begin{equation}
\ket{\Phi_{j_\ell}}=\left(\frac{a_++a_-}{\sqrt{2}}\right)\ket{\Psi_{+,j_\ell}}+\left(\frac{a_+-a_-}{\sqrt{2}}\right)\ket{\Psi_{-,j_\ell}}.
\end{equation}
It is also interesting for the following to express $\ket{\Psi_{\pm,j_\ell}}$ in cylindrical coordinates ($\overrightarrow{e}_r,\overrightarrow{e}_\theta$):
\begin{equation}
\ket{\Psi_{+,j_\ell}}=e^{ij_\ell\theta}\overrightarrow{e}_r,\, \ket{\Psi_{-,j_\ell}}=ie^{ij_\ell\theta}\overrightarrow{e}_\theta.
\label{EigenVectorTAM_polar_coordinates}
\end{equation}
\subsection{Total angular momentum eigenvectors: an orthonormal basis for representing the electric field}
First, one has to emphasize that analysing only the transverse part of the field $\overrightarrow{\tilde{E}}^\perp(r,\theta)$ is sufficient at this stage. This is because, first, the longitudinal part of the field $\overrightarrow{\tilde{E}}_z$ can be retrieved using the divergence constraint (Eq.\,\ref{Eq_Divergence}) and secondly, because this component is generally speaking extremely weak in fibers. For simplicity, we will then use the shorthand notation $\overrightarrow{\tilde{E}}$ for the transverse part. Moreover, note that all derivations made below apply in the frequency domain. However, we will omit when not necessary the notation specifying that the field is expressed in the frequency domain.\\
Since the electric field is intrinsically angularly $2\pi$ periodic, both circularly polarized field components can be decomposed as a Fourier series in polar angle:
\begin{equation}
\overrightarrow{E}(r,\theta)=\sum_\ell{a_{+,\ell}(r)e^{i\ell\theta}\ket{\sigma_+}+a_{-,\ell}(r)e^{i\ell\theta}\ket{\sigma_-}}.
\label{FieldCircularBasis}
\end{equation}
In cylindrical coordinates, it writes
\begin{eqnarray}
\begin{aligned}
\overrightarrow{E}(r,\theta)&=\sum_\ell{\frac{a_{+,\ell}(r)}{\sqrt{2}}e^{i(\ell+1)\theta}\left(\overrightarrow{e}_r+i\overrightarrow{e}_\theta\right)}&\\
&+\sum_\ell{\frac{a_{-,\ell}(r)}{\sqrt{2}}e^{i(\ell-1)\theta}\left(\overrightarrow{e}_r-i\overrightarrow{e}_\theta\right)},&
\end{aligned}
\end{eqnarray}
which can be recast as
\begin{equation}
\overrightarrow{E}(r,\theta)=\sum_{j_\ell}{A_{+,j_\ell}(r)\ket{\Psi_{+,j_\ell}}+A_{-,j_\ell}(r)\ket{\Psi_{-,j_\ell}}},
\end{equation}
with $A_{\pm,j_\ell}(r)=\frac{a_{+,\ell}(r)\pm a_{-,\ell+2}(r)}{\sqrt{2}}$.
The eigenvectors $\ket{\Psi_{\pm,j_\ell}}$ hence form an orthonormal basis, any electric field being represented as a unique weighted sum of TAM eigenvectors.
\subsection{Spin-orbit coupling}
Spin-orbit coupling \cite{BliokhReview} can be highlighted by looking at the action of the operator $\overrightarrow{S}$ on a circularly polarized field embedding a well defined orbital angular momentum. Let us consider a circularly polarized electric field $\overrightarrow{E}_{(\ket{\sigma_\pm},\ell)}$ embedding an orbital angular momentum $\ell$.
\begin{equation}
\overrightarrow{E}_{(\ket{\sigma_\pm},\ell)}=A(r)e^{i\ell\theta}\ket{\sigma_\pm.}
\end{equation}
In cylindrical coordinates, the electric field components ($E_r$, $E_\theta)$ write
\begin{eqnarray}
\begin{aligned}
&E_r\left(\ket{\sigma_\pm},\ell\right)=A(r)e^{i\left(\ell\pm1\right)\theta},&\\
&E_\theta\left(\ket{\sigma_\pm},\ell\right)=\pm iA(r)e^{i\left(\ell\pm1\right)\theta}.&
\end{aligned}
\label{Eq2}
\end{eqnarray}
Injecting Eq.\,\ref{Eq2} in Eq.\,\ref{Eq1}, one has
\begin{eqnarray}
\overrightarrow{S}\left[\overrightarrow{E}_{(\ket{\sigma_\pm},\ell)}\right]=
\begin{pmatrix}
\partial_r^2fA(r)+\partial_rf\partial_rA(r)\\
\frac{i\left(\ell\pm1\right)}{r}\partial_rfA(r)
\end{pmatrix}
e^{i\left(\ell\pm1\right)\theta}.
\label{Eq3}
\end{eqnarray}
The vector $\overrightarrow{S}$ can be finally projected back in the circular polarization basis:
\begin{eqnarray}
\begin{aligned}
S_{\ket{\sigma_+}}=\frac{S_r-iS_\theta}{\sqrt{2}}e^{-i\theta}\\
S_{\ket{\sigma_-}}=\frac{S_r+iS_\theta}{\sqrt{2}}e^{i\theta},
\end{aligned}
\end{eqnarray}
where $S_r$ (resp. $S_\theta$) is the component along $\overrightarrow{e}_r$ (resp. $\overrightarrow{e}_\theta$) of $\overrightarrow{S}$ given in Eq.\,\ref{Eq3}.
One then notices that the term $\overrightarrow{S}$ induces a spin-orbit coupling through the refractive index gradient. More particularly, for a circularly polarized electric field ($\ket{\sigma_\pm}$), the component of opposite helicity (i.e., along $\ket{\sigma_\mp}$) of $\overrightarrow{S}$ is
\begin{equation}
S_{\ket{\sigma_\mp}}=\frac{\partial^2_rfA(r)+\partial_rf\partial_rA(r)\mp\frac{\ell\pm1}{r}\partial_rfA(r)}{\sqrt{2}}e^{i\left(\ell\pm2\right)\theta}.
\end{equation}
One can readily deduce the following selection rules that apply in circularly symmetric media:
\begin{eqnarray}
\boxed{
\begin{aligned}
\left(\ket{\sigma_+},\ell\right)\overset{\overrightarrow{S}}{\rightarrow}\left(\ket{\sigma_-},\ell+2\right),\\
\left(\ket{\sigma_-},\ell\right)\overset{\overrightarrow{S}}{\rightarrow}\left(\ket{\sigma_+},\ell-2\right),\\
\end{aligned}
}
\end{eqnarray}
The operator $\overrightarrow{S}$ then does not conserve neither the spin nor the orbital angular momentum on their own but nevertheless conserves the projection along the $z$ axis of the TAM operator $\mathcal{J}=\mathcal{L}+\mathcal{S}$ as we discuss now.
\begin{figure}[t!]
\begin{center}
\includegraphics[width=8cm,keepaspectratio]{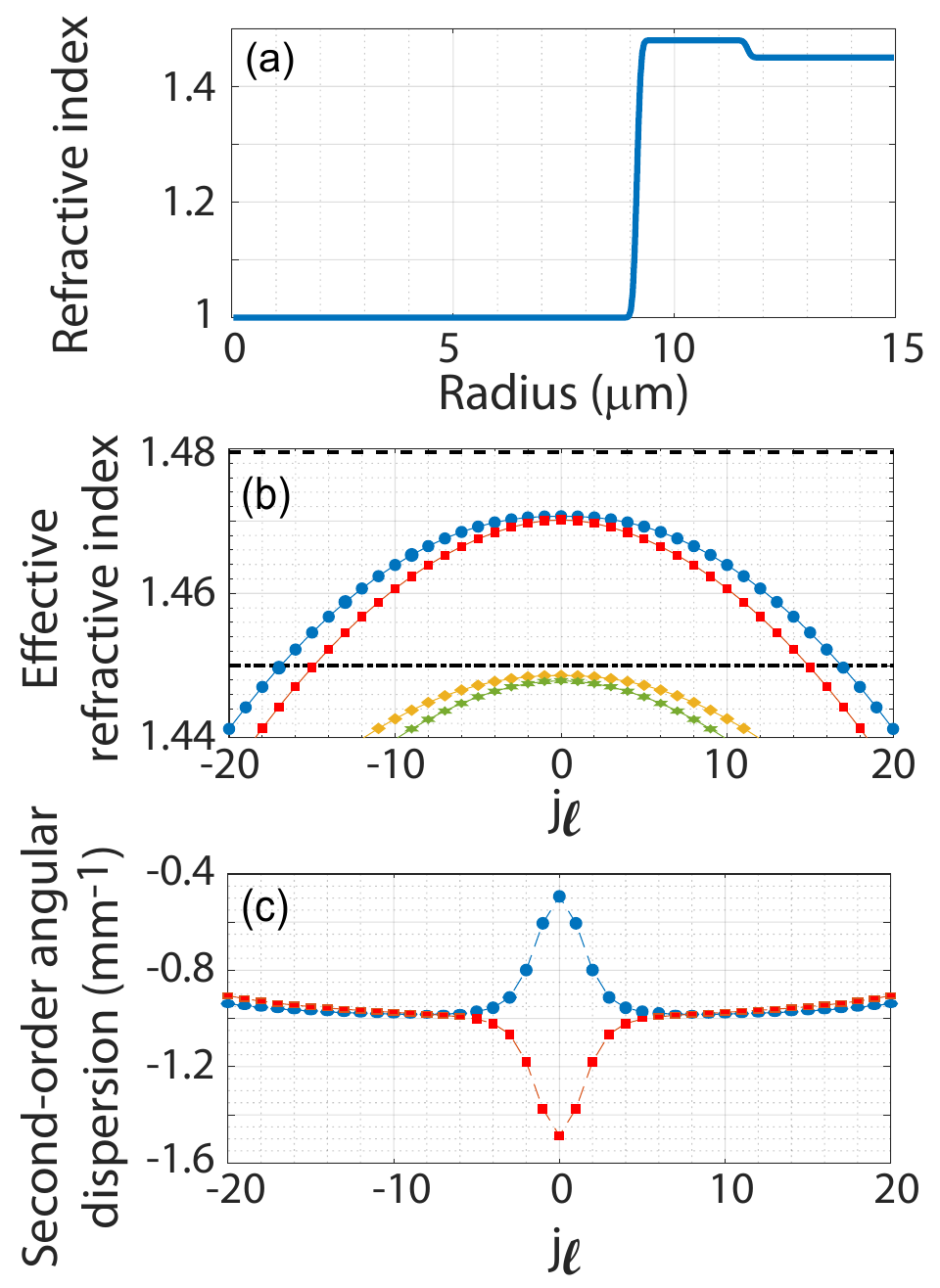}
	\caption{Refractive index profile of the air-core fiber used in the present example (a) and propagation constants as a function of the TAM $j_\ell$ of the vectorial modes (b). Only one group of radial modes is guided within the ring (blue squares and red circles) while the second (and higher) order radial modes are not (orange diamonds and green stars). The modes depicted as blue squares and red circles differ by their distinct polarization patterns. Panel (c) displays the second-order derivative of the guided modes propagation constants with respect to $j_\ell$.}
 	\label{Fig1}
 \end{center}
\end{figure}
\begin{figure*}
\begin{center}
\includegraphics[width=18cm,keepaspectratio]{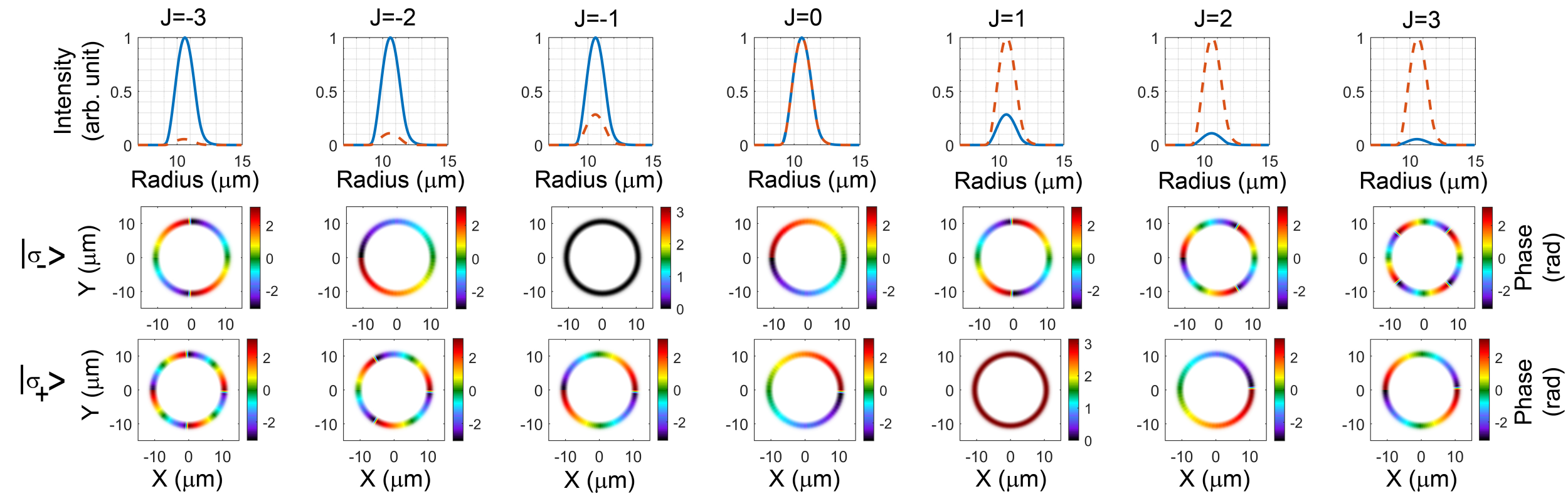}
\caption{Electric field transverse profiles of the 7 first modes of the first group. Each column corresponds to a mode of well-defined TAM. The first row depicts the radial intensity profile of the circular left (solid blue) and right (dashed red) polarization component of the mode. The middle (resp. bottom) row depicts the spatial phase profile of the circular left (resp. right) component of the mode.}
\label{Groupe1}
\end{center}
\end{figure*}
\subsection{Total angular momentum conservation in circularly symmetric media}
The conservation of TAM can be highlighted by looking at the action of the propagation operator $\overrightarrow{\Box}$ (Eq.\,\ref{BoxOperator}) on an electromagnetic field embedding a well defined TAM $j_\ell$. In this regard, it is sufficient to express the propagation equation for the two orthonormal eigenvectors of the TAM (Eq.\,\ref{basisSet}). As described previously (Eq. \ref{EqPropagationVectorielle}), the propagation equation of the electric field spectrum $\overrightarrow{\tilde{E}}$ in a non-magnetic, inhomogeneous, isotropic and linear medium reads
\begin{equation}
\overrightarrow{\Box}\overrightarrow{\tilde{E}}=\overrightarrow{0}.
\end{equation}
Let us now consider the following two fields $\overrightarrow{\tilde{E}}_{\pm,j_\ell}$
\begin{equation}
\overrightarrow{\tilde{E}}_{\pm,j_\ell}=A_{\pm,j_\ell}(r)\ket{\Psi_{\pm,j_\ell}}e^{iK_zz},
\end{equation}
that own the same TAM $j_\ell$. Injecting them into Eq.\,\ref{EqPropagationVectorielle}, one obtains
\begin{eqnarray}
\begin{aligned}
\nabla^2\overrightarrow{\tilde{E}}_{\pm,j_\ell}+\frac{n(r)^2\omega^2}{c^2}\overrightarrow{\tilde{E}}_{\pm,j_\ell}&=\mathcal{D}_{+,j_\ell}\left[A_{\pm,j_\ell}(r)\right]e^{iK_zz}\ket{\Psi_{\pm,j_\ell}}&\\
&+\mathcal{D}_{-,j_\ell}\left[A_{\pm,j_\ell}(r)\right]e^{iK_zz}\ket{\Psi_{\mp,j_\ell}},&
\end{aligned}
\end{eqnarray}
with
\begin{eqnarray}
\begin{aligned}
&\mathcal{D}_{+,j_\ell}=\left(\partial^2_r+\frac{1}{r}\partial_r-\frac{j_\ell^2+1}{r^2}+\frac{n(r)^2\omega^2}{c^2}-K_z^2\right),&\\
&\mathcal{D}_{-,j_\ell}=\frac{2j_\ell}{r^2}.&
\label{EqD}
\end{aligned}
\end{eqnarray}
and, in the case of circularly symmetric inhomogeneous media,
\begin{eqnarray}
\begin{aligned}
\overrightarrow{S}[\overrightarrow{\tilde{E}}_{+,j_\ell}]&=S_{+,j_\ell}[A_{\pm,j_\ell}(r)]e^{iK_zz}\ket{\Psi_{\pm,j_\ell}}&\\
&+S_{-,j_\ell}[A_{\pm,j_\ell}(r)]e^{iK_zz}\ket{\Psi_{\mp,j_\ell}},&\\
\overrightarrow{S}[\overrightarrow{\tilde{E}}_{-,j_\ell}]&=\overrightarrow{0}.&
\label{EqS}
\end{aligned}
\end{eqnarray}
with
\begin{eqnarray}
\begin{aligned}
&\mathcal{S}_{+,j_\ell}=\partial^2_rf+\partial_rf\partial_r,&\\
&\mathcal{S}_{-,j_\ell}=\frac{\ell+1}{r}\partial_rf.&
\end{aligned}
\end{eqnarray}
The action of the propagation operator hence couples only states owing the same TAM, confirming thereby that \textit{the propagation operator $\overrightarrow{\Box}$ preserves the TAM of light}. Other said, the electromagnetic fields that are invariant at a given absolute phase over the propagation (i.e., optical modes) in circularly symmetric inhomogeneous media are also necessarily eigenvectors of TAM. This remark then implies that optical modes necessarily take the following functional form
\begin{equation}
\overrightarrow{E}_\textrm{mode}(r,\theta,z)=\left(A_{+,j_\ell}(r)\ket{\Psi_{+,j_\ell}}+A_{-,j_\ell}(r)\ket{\Psi_{-,j_\ell}}\right)e^{iK_zz},
\end{equation}
which can be recast in cylindrical coordinates as
\begin{equation}
\overrightarrow{E}_\textrm{mode}(r,\theta,z)=e^{ij_\ell\theta}\left[A_{r,\ell}(r)\overrightarrow{e}_{r}+iA_{\theta,\ell}(r)\overrightarrow{e}_{\theta}\right]e^{iK_zz}.
\label{ModeCylindricalCoordinates}
\end{equation}
Before presenting how the radial part of the modes can be evaluated, it is interesting to have a close look on the structure of the matrix coupling $\ket{\Psi_{+,j_\ell}}$ and $\ket{\Psi_{-,j_\ell}}$. Schematically, according to Eqs.\,(\ref{EqD}-\ref{EqS}), the coupling matrix $C_{j_\ell}$ has the following form:
\begin{equation}
C_{j_\ell}=
\begin{pmatrix}
d_++S_+-\frac{j_\ell^2}{r^2}&j_\ell d_-\\
j_\ell\left(d_-+s_-\right)&d_+-\frac{j_\ell^2}{r^2}
\end{pmatrix},
\label{CouplingMatrix}
\end{equation}
where $d_+=\partial^2_r+\frac{1}{r}\partial_r-\frac{1}{r^2}+\frac{n(r)^2\omega^2}{c^2}$, $d_-=\frac{2}{r^2}$, and $s_-=\frac{1}{r}\partial_rf$. Calculating the optical modes amounts to solve the equation system $C_{j_\ell}\ket{\phi_{j_\ell}}=K_z^2\ket{\phi_{j_\ell}}$, where $\ket{\phi_{j_\ell}}$ is an eigenvector of $C_{j_\ell}$ (and consequently a mode of the waveguide) and $K_z^2$ is its associated eigenvalue that corresponds to the square of the modal propagation constant. First, it is clear that the determinant of $C_{j_\ell}$ only depends on $j_\ell^2$. As a consequence, the modal propagation constants will not depend on the sign of $j_\ell$. This essential degeneracy, linked to the system symmetry, reflects that there is no preferential rotation direction for the field. Moreover, for not too strong refractive index gradient, one has $\mathcal{S}\ll\mathcal{D}$. Looking at Eq.\,\ref{CouplingMatrix} for $j_\ell\ne0$, in the limiting case $\mathcal{S}\to 0$, optical modes will tend to the following functional forms:
\begin{eqnarray}
\begin{aligned}
\overrightarrow{E}_\textrm{mode}(r,\theta,z)&\simeq A(r)e^{ij_\ell\theta}\left[\overrightarrow{e}_{r}\pm i\overrightarrow{e}_{\theta}\right]e^{iK_zz},&\\
&\simeq A(r)e^{i\left(j_\ell\mp1\right)\theta}e^{iK_zz}\ket{\sigma_{\pm}}.&
\end{aligned}
\end{eqnarray}
The modes $j_\ell\ne0$ are then in good approximation circularly polarized owing a well defined orbital angular momentum. Note that this last remark only holds for fiber exhibiting a not too strong refractive index gradient, such as conventional fibers. This is not the case, for instance, for ring-core fibers, as we will discuss a bit later. On the contrary, when $j_\ell=0$, $\ket{\Psi_{+,0}}$ and $\ket{\Psi_{-,0}}$ are not coupled anymore. They are thus eigenvectors of the propagation equation but do not share the same eigenvalue, excepted in the bulk case. The modes for $j_\ell=0$ will then be either radially (i.e., along $\overrightarrow{e}_r$) or azimuthally (i.e., along $\overrightarrow{e}_\theta$) polarized:
\begin{eqnarray}
\begin{aligned}
&\overrightarrow{E}_{+,j_\ell=0}(r,\theta,z)=A_r(r)e^{iK_rz}\overrightarrow{e}_r,&\\
&\overrightarrow{E}_{-,j_\ell=0}(r,\theta,z)=A_\theta(r)e^{iK_\theta z}\overrightarrow{e}_\theta,&
\label{modeJ0}
\end{aligned}
\end{eqnarray}
with $K_r\ne K_\theta$. The next section is now devoted to describe how the modal transverse profile can be calculated efficiently.
\subsection{Determination of the radial distribution of the modes}
As discussed before, optical modes of circularly symmetric waveguides are necessarily eigenvectors of the TAM operator so that their general functional forms follow Eq.\,\ref{ModeCylindricalCoordinates}. One has to emphasize here that the refractive index derivatives appearing in $C_{j_\ell}$ has to be understood in the sense of distributions that make possible to differentiate functions whose derivatives do not exist in the classical sense. This point is of particular importance when one deals with step index fibers whose refractive index profiles exhibit discontinuities. As far as the radial distribution of the modes is concerned, it can be numerically evaluated for any refractive index distribution by using a convenient representation basis for the radial dimension. Note again that this procedure has to be repeated, if needed, for every frequencies independently. As it was done in our previous scalar algorithm \cite{BejotPRE}, we decided here to use a radial basis set composed of zero order Bessel functions $J_0(k_\perp r)$. The main advantage of such a choice is that it is extremely easy to go back and forth between a direct space representation of the field and its representation in the Bessel basis (through a fast Hankel transform with its related unitary transformation matrix $H$ \cite{HankelTransformAlgo}). In the present case, the basis set used for the calculation is then composed of vector field functions $\mathcal{\textbf{B}}$ that follow the general functional form given in Eq.\,\ref{ModeCylindricalCoordinates}
\begin{equation}
\mathcal{\textbf{B}}=e^{ij_\ell\theta}\left[J_0\left(k_\perp r\right)\overrightarrow{e}_r+iJ_0\left(k'_\perp r\right)\overrightarrow{e}_\theta\right].
\end{equation}
Numerically speaking, the theoretically infinite and dense Bessel basis becomes finite and discrete. This is because a finite and discrete number of radial points is used. As a result, sampling the radial part of each field polarization component with $N_r$ spatial points implies the use of a Bessel radial basis containing $N_r$ vectors. The numerical procedure for calculating the optical modes consequently amounts, for any given TAM $j_\ell$, to diagonalize $C_{j_\ell}$ expressed in the basis $\mathcal{\textbf{B}}$ and whose dimension is $4N_r^2$ (the factor 4 coming from the fact that the subspace associated to a given $j_\ell$ is bi-dimensional). The diagonalization returns $2N_r$ orthonormal eigenvectors $v_{j_\ell}$ that form an orthonormal basis $\mathcal{B}_{\textrm{eig}}(j_\ell)$. The matrix $V_{j_\ell}$, filled with the coordinates of the eigenvectors $v_{j_\ell}$ in the basis $\mathcal{\textbf{B}}$, is then the transformation matrix from $\mathcal{\textbf{B}}$ to $\mathcal{B}_{\textrm{eig}}(j_\ell)$. As a result, once an electric field is decomposed in the basis $\mathcal{\textbf{B}}$ (which corresponds to the Hankel transform of the radial profiles of both polarization components), it is extremely easy to describe it in the modal basis. In the following section, we will see in more details how this can be done numerically in an efficient way.
\subsection{Fast modal transform}
\label{SectionModalTransform}
Let us assume that the electric field $\overrightarrow{E}$ to be decomposed in the eigenbasis is initially written in the direct space (in polar coordinates) and in the circular polarization basis
\begin{equation}
    \overrightarrow{E}(r,\theta)=a_+(r,\theta)\ket{\sigma_+}+a_-(r,\theta)\ket{\sigma_-}.
\end{equation}
Note that, numerically speaking, the coordinates ($r$ and $\theta$) are sampled with a number of points $N_r$ and $N_\theta$ respectively. For simplicity, the field is written as a matrix of dimension $2N_r\times N_\theta$, where the first (resp. last) $N_r$ rows correspond to the left (resp. right) circular polarization component. Note that the angular sampling implies that only a finite number of TAM values is considered ($j_\ell \in \left[-\frac{N_\theta}{2},\frac{N_\theta}{2}-1\right]$). The field can be easily recast so as to be written along the vectors $\overrightarrow{e}_r$ and $\overrightarrow{e}_\theta$ with the following transformation $T$
\begin{equation}
    \overrightarrow{E}(r,\theta)=a_r(r,\theta)\overrightarrow{e}_r+ia_\theta(r,\theta)\overrightarrow{e}_\theta,
\end{equation}
with
\begin{equation}
a_r(r,\theta)=\frac{e^{i\theta}a_++e^{-i\theta}a_-}{\sqrt{2}},\, a_\theta(r,\theta)=\frac{e^{i\theta}a_+-e^{-i\theta}a_-}{\sqrt{2}}.
\end{equation}
\begin{figure*}[t!]
\begin{center}
\includegraphics[width=18cm,keepaspectratio]{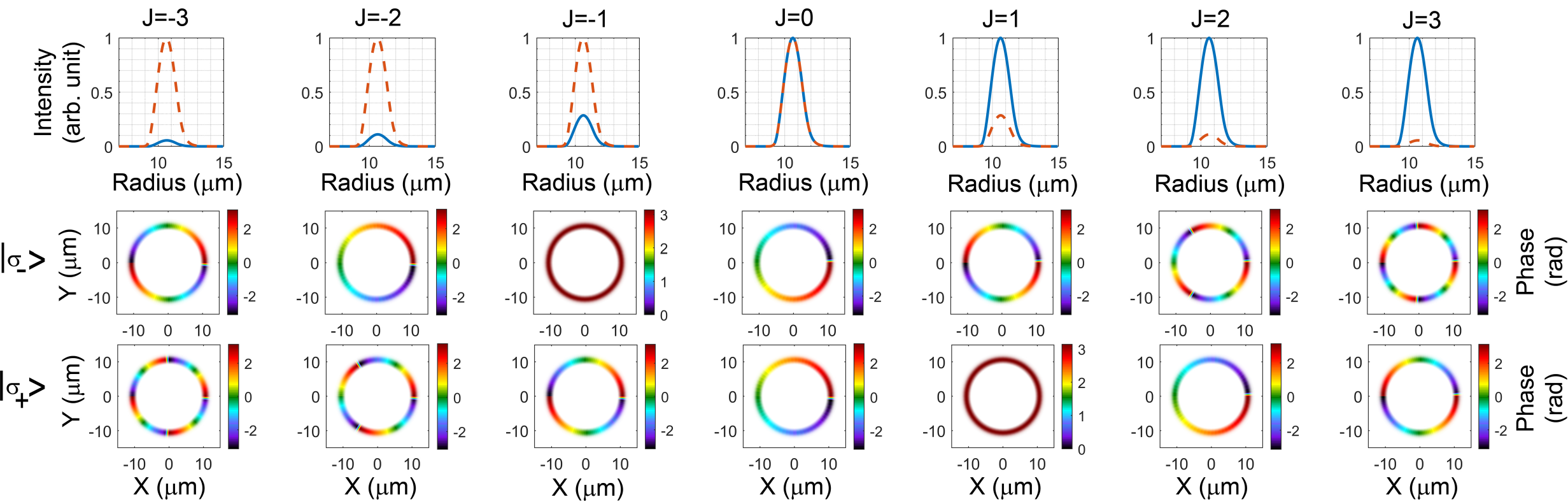}
\caption{Electric field transverse profiles of the 7 first modes of the second group. Each column corresponds to a mode of well-defined TAM. The first row depicts the radial intensity profile of the circular left (solid blue) and right (dashed red) polarization component of the mode. The middle (resp. bottom) row depicts the spatial phase profile of the circular left (resp. right) component of the mode.}
\label{Groupe2}
\end{center}
\end{figure*}
One can now decompose the two field components as an angular Fourier series
\begin{eqnarray}
\begin{aligned}
&\overrightarrow{E}=\sum_{j_\ell}{e^{j_\ell\theta}\left[a_{r,j_\ell}(r)\overrightarrow{e}_r+ia_{\theta,j_\ell}(r)\overrightarrow{e}_\theta\right]}&\\
&\overrightarrow{E}=\sum_{j_\ell}{a_{r,j_\ell}(r)\ket{\Psi_{+,j_\ell}}+a_{\theta,j_\ell}(r)\ket{\Psi_{-,j_\ell}}}&,
\end{aligned}
\end{eqnarray}
where the coefficients $a_{r,j_\ell}$ and $a_{\theta,j_\ell}$ are easily obtained by making a fast Fourier transform $FFT_\theta$ of the field matrix along the second dimension (i.e., along $\theta$). After this operation, the field is represented, at each $r$, in the basis of the TAM operator (Eq.\,\ref{EigenVectorTAM_polar_coordinates}). Numerically speaking, the $J^{\textrm{th}}$ column (of length $2N_r$) of the matrix representing the field is then now $\left[a_{r,j_\ell(J)},a_{\theta,j_\ell(J)}\right]$. If one now wants to express the field in the modal basis, the last remaining operation is to multiply each column by the matrix $M_{j_\ell}=V_{j_\ell}H$, where $H$ is the Hankel transform unitary matrix \cite{HankelTransformAlgo}. Note that the matrix $M_{j_\ell}$ to use is not the same for each column (i.e., for each $j_\ell$). Numerically speaking, it is then advantageous to reshape at this point the field matrix as a column vector (of length $2N_rN_\theta$) and define a sparse block diagonal matrix $M$ (of dimension $4N_r^2N^2_\theta$) that is defined as
\begin{equation}
M=
\begin{pmatrix}
M_{j_1} & (0) & (0)\\
(0) & \ddots & (0)\\
(0) & (0) & M_{j_{N_\theta}}
\end{pmatrix}
\label{MatrixMj}
\end{equation}
Doing so avoids a numerically heavy loop over $j_\ell$ and allows to express the field in the modal basis for every $j_\ell$ at once. After the operation, the field is expressed in the modal basis and can be reshaped back into a $2N_r\times N_\theta$ matrix. The global operation for going from the real space to the modal basis finally reads
\begin{equation}
\bar{E}\left(p,j_\ell\right)=\mathcal{R}_2^{-1}M\mathcal{R}_2\left(FFT_\theta\left[T\overrightarrow{E}(r,\theta)\right]\right),
\label{ModalTransform}
\end{equation}
where $\mathcal{R}_2$ is the numerical operation for transforming the 2D matrix into a column vector and $\bar{E}\left(p,j_\ell\right)$ is the electric field expressed in the modal basis. Equation \ref{ModalTransform} constitutes the fast modal transform that allows to go back and forth from the real space to the modal space.
Note that, we considered here that the initial field was monochromatic. When dealing with pulsed fields, the transformation described above as to be performed in the frequency domain since the transformation matrix (Eq.\,\ref{MatrixMj}) obtained during the diagonalization is now frequency dependent and denoted hereafter $M_\omega$. In this case, the modal transform of a field $\overrightarrow{E}(r,\theta,t)$ reads
\begin{equation}
\bar{E}\left(p,j_\ell,\omega\right)=\mathcal{R}_3^{-1}M_\omega\mathcal{R}_3\left(FFT_\theta\left[FFT_t\left(T\overrightarrow{E}(r,\theta,t)\right)\right]\right),
\label{ModalTransformT}
\end{equation}
where $FFT_t$ is the one-dimensional fast Fourier transform operated along the time dimension, $\mathcal{R}_3$ stands for the numerical operation transforming the 3D matrix representing the field into a column vector, and $M_\omega$ is the following sparse block diagonal transformation matrix (of dimension $4N_r^2N^2_\theta N_t^2$, where $N_t$ is the number of sampling points used in the time domain):
\begin{equation}
M_\omega=
\begin{pmatrix}
M_{\omega_1} & (0) & (0)\\
(0) & \ddots & (0)\\
(0) & (0) & M_{\omega_{N_t}}
\end{pmatrix}
\end{equation}
The modal transform described above can be used for projecting any vector field distributions in the modal basis. As we will see later, it will be particularly useful for solving the nonlinear unidirectional pulse propagation equation.
\subsection{Numerical example}
\label{RingCoreMode}
For the present example, we consider an air-core fiber whose (circularly symmetric) refractive index profile is depicted Fig.\,\ref{Fig1}(a). Such kind of fiber is now well-known for allowing the propagation of modes embedding orbital angular momenta with extremely low cross-talks over long distances. We consider for this example a wavelength of $1.035\mu$m. We used 250 (resp. 64) points for sampling the radial (resp. angular) coordinate, meaning that a total of 32000 modes (guided in the ring-core or not) have been calculated. As shown in Fig.\,\ref{Fig1}(b), among all modes calculated, the considered fiber supports a total of 62 modes guided within the ring-core, all belonging to the same group of radial modes. The guided modes can be divided into two distinct groups of modes, whose (spatially-dependent) polarization patterns are orthogonal. Each mode belonging to a particular group of modes has a distinct TAM $j_\ell$. Note that modes of the same group but with opposite TAM own exactly the same effective refractive index. Figure \ref{Groupe1} (resp. Fig.\,\ref{Groupe2}) depicts the seven first guided modes of the first (resp. second) polarization group. As shown, all modes share almost the same radial distribution. Moreover, they are characterized by their strong vectorial nature, i.e., their spatially-depend polarization pattern. However the polarization of the modes gradually approaches to a pure circular polarization as the TAM value increases. This can be seen by noticing that the amplitude of one polarization component, relatively to the other, becomes weaker and weaker as the TAM $|j_\ell|$ increases. As far as the phase spatial distribution is concerned (middle and bottom rows), one retrieves the expected signature of fields owing a well-defined TAM (Eq.\,\ref{EigenVectorTAM}). After having presented the algorithm for calculating the modes, their associated propagation constants and the modal transform, one can now describe how to solve the unidirectional pulse propagation equation.
\section{Numerical algorithm for solving the nonlinear unidirectional pulse propagation equation}
\label{UPPESolving}
\subsection{Nonlinear unidirectional pulse propagation equation}
When nonlinear effects are taken into account, the electric displacement field $\overrightarrow{D}$ reads in the frequency domain
\begin{equation}
\overrightarrow{\tilde{D}}=\epsilon_0\left(1+\chi^{(1)}\right)\overrightarrow{\tilde{E}}+\overrightarrow{\tilde{P}}_{\textrm{NL}},
\end{equation}
where $\overrightarrow{P}_{\textrm{NL}}$ is the nonlinear polarization.
The divergence of the electric field then now verifies
\begin{equation}
\overrightarrow{\nabla}\cdot\overrightarrow{\tilde{E}}=-\left[\overrightarrow{\nabla}\left(\textrm{log}\,n^2\right)\cdot\overrightarrow{\tilde{E}}+\frac{\overrightarrow{\nabla}\cdot\overrightarrow{\tilde{P}}_{\textrm{NL}}}{\epsilon_0n^2}\right],
\end{equation}
where $n=\sqrt{1+\chi^{(1)}}$ is the linear refractive index. As a result, the nonlinear propagation equation now reads in the frequency domain
\begin{equation}
\overrightarrow{\Box}\overrightarrow{\tilde{E}}=-\frac{\omega^2\overrightarrow{\tilde{P}}_{\textrm{NL}}}{\epsilon_0c^2}-\frac{\overrightarrow{\nabla}\left(\overrightarrow{\nabla}\cdot\overrightarrow{\tilde{P}}_{\textrm{NL}}\right)}{\epsilon_0n^2},
\end{equation}
with $\overrightarrow{\Box}$ the operator defined in Eq.\,\ref{BoxOperator}. If one projects this equation on the modal space (i.e., on the basis composed of the vectorial modes of the linear propagation equation) and neglecting the backward field, one obtains the nonlinear UPPE equation
\begin{equation}
\partial_z\bar{E}=iK_z\bar{E}+\frac{i}{2\epsilon_0K_z}\left(\frac{\omega^2\bar{P}_{\textrm{NL}}}{c^2}+\overline{\left[\frac{\overrightarrow{\nabla}\left(\overrightarrow{\nabla}\cdot\overrightarrow{P}_{\textrm{NL}}\right)}{n^2}\right]}\right),
\label{UPPE_Equation}
\end{equation}
where $\bar{A}$ stands for the decomposition of a three-dimensional vector field $\overrightarrow{A}$ on the modal basis and $K_z$ is the associated propagation constant. Note here that Eq.\,\ref{UPPE_Equation} takes into account both the transverse field and its associated longitudinal component (i.e., the field component polarized along $z$). For a more detailed derivation of the nonlinear unidirectional pulse propagation equation, we invite the reader to refer to \cite{KolesikUPPE}. In the context of propagation within an optical fiber, a fair approximation is to neglect the last term in Eq.\,\ref{UPPE_Equation}, and also the longitudinal component of the field since this part remains generally small as compared to its transverse part. As a result, its contribution to the nonlinear polarization will be neglected. As a result, we will now consider that the field is in fair approximation purely transverse (i.e., contained in the $\left[r,\theta\right]$ plane). However, one has to keep in mind that these approximations break when the field embeds modes that have propagation constants far smaller than the propagation constant in bulk ($K_z\ll\frac{n\omega}{c}$).
\begin{figure}
    \begin{center}
\includegraphics[width=9cm,keepaspectratio]{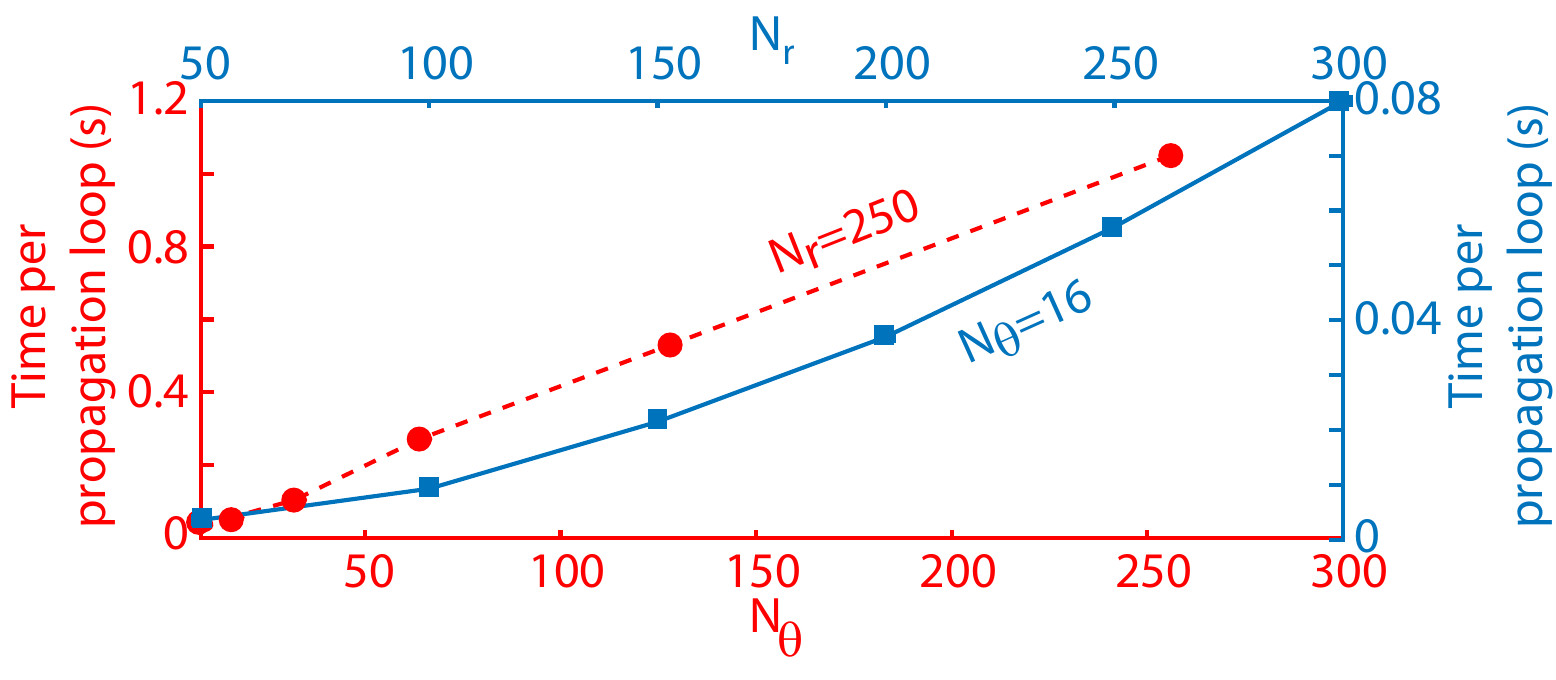}
	\caption{Computation time per propagation loop as a function of the number of radial (solid blue) and angular (dashed red) points. The solid blue (resp. dashed red) curve has been calculated with $N_\theta$=16 (resp. $N_r$=250).}
 	\label{Fig4}
 \end{center}
\end{figure}
\subsection{Complex envelope equation}
Most of the time, it is more convenient to deal with a propagation equation acting on the complex envelope of the field. This is done by defining the complex field $\overrightarrow{\varepsilon}$ as
\begin{equation}
\overrightarrow{E}=\frac{\overrightarrow{\varepsilon}+\overrightarrow{\varepsilon^*}}{2}.
\end{equation}
For convenience, we also introduce the vector field
$\overrightarrow{\xi}$ as
\begin{equation}
\overrightarrow{\xi}=\sqrt{\frac{2}{\epsilon_0cn}}\overrightarrow{\varepsilon},
\end{equation}
so that the total intensity $I$ reads
\begin{equation}
I=\overrightarrow{\xi}\cdot\overrightarrow{\xi^*}.
\end{equation}
In the context of optical fiber propagation, we consider that the medium is in good approximation isotropic, at least, in a microscopic point of view. In this regard, the number of independent elements of the third-order nonlinear susceptibility tensor $\chi^{(3)}$ reduces to one.We can consequently define an unique nonlinear refractive index $n_2$ as
\begin{equation}
n_2=\frac{3\chi^{(3)}_{xxxx}}{4\epsilon_0cn_0^2},
\end{equation}
where $n_0$ is the bulk refractive index taken at the central frequency of the considered field. Neglecting the term responsible for third harmonic generation and considering at this point a pure electronic instantaneous Kerr effect, the nonlinear polarization reads in the circular polarization basis
\begin{equation}
\overrightarrow{P}_{\textrm{NL}}=\frac{2n_0n_2}{3}
\begin{pmatrix}
\left[|\xi_+|^2+2|\xi_-|^2\right]\xi_+\\
\left[2|\xi_+|^2+|\xi_-|^2\right]\xi_-
\end{pmatrix}
,
\label{PNL}
\end{equation}
where $\xi_+$ (resp. $\xi_-$) is the left (resp. right) circular component of $\overrightarrow{\xi}$. Note that, if needed, the Raman-induced contribution to the nonlinear polarization can also be added. As one can remark, the nonlinear polarization term is easily calculated in the direct space (and in the circular polarization basis). Finally, it is of common use to consider a sliding time origin (corresponding, at any propagation distance $z$, to the arrival time of a pulse propagating at a group velocity $v_{g_0}$). The complex unidirectional pulse propagation finally reads in the modal basis
\begin{equation}
\partial_z\bar{\xi}=i\left(K_z-\frac{\omega}{v_{g_0}}\right)\bar{\xi}+i\frac{\omega^2}{K_zc^2}\bar{P}_{\textrm{NL}}.
\label{EqUPPEFinal}
\end{equation}
In the next section, we will see how Eq.\,\ref{EqUPPEFinal} can be numerically resolved in an efficient way using the modal transform presented in section\,\ref{SectionModalTransform}.
\subsection{Numerical algorithm}
The starting point for the numerical algorithm presented here is to note that the linear term can be easily evaluated in the modal space while it is far more convenient to calculate the nonlinear one in the direct space. As a result, the strategy for solving Eq.\,\ref{EqUPPEFinal} is to use a split-step algorithm during which both terms (linear and nonlinear) are evaluated in a different representation space. In this regard, one can see the present method as a split-step modal algorithm, in analogy with the well-known split-step Fourier and split-step Hankel-Fourier algorithms. The proposed algorithm is then extremely close to the one we presented for solving the scalar version of UPPE \cite{BejotPRE}. The key point of the algorithm lies in the construction of the modal transform, described in section \ref{SectionModalTransform} (Eq.\,\ref{ModalTransformT}), for being able to efficiently go from the direct space to the modal space (and \textit{vice versa}). Knowing the expression of a field $\overrightarrow{\xi}(r,\theta,z,t)$ at a given $z$, one can calculate the field at a close distance $z+dz$ as follow. First, one evaluates its expression $\bar{\xi}$ in the modal basis using Eq.\,\ref{ModalTransformT}. Then, a linear propagation step over $dz/2$ is performed so that
\begin{equation}
    \bar{\xi}\rightarrow e^{i\left(K_z-\frac{\omega}{v_{g_0}}\right)\frac{dz}{2}}\bar{\xi}.
\end{equation}
After this, the nonlinear polarization is evaluated in the direct space. This is done by evaluating the new field $\bar{\xi}$ in the direct space. Once calculated in the direct space, the nonlinear polarization is expressed in the modal basis with the modal transform so as to solve the second term in Eq.\,\ref{EqUPPEFinal} by a fourth-order Runge-Kutta algorithm. Finally, a second linear step over $dz/2$ is performed in the modal basis. Doing so avoids to deal with multiple summations of overlap integrals as it is usually done in simulations solving the multimode-generalized nonlinear Schrodinger equation. In order to assess the efficiency of the algorithm, we evaluated how the computation time increases with the number of modes involved in the code [which is equal to $2N_rN_\theta$, where $N_r$ (resp. $N_\theta$) is the number of points in the radial (resp. angular) dimension]. For this, we timed one hundred nonlinear propagation loops as a function of $N_r$ and $N_\theta$ in the monochromatic case ($N_t$=1). The calculations have been performed in a Dell Optiplex 5260 computer (32 Go RAM, processor Intel $I7$ 8$^{\textrm{th}}$ generation). The results are summarized in Fig.\,\ref{Fig4}. As shown, the propagation algorithm complexity is approximately $\mathcal{O}\left(Nr^2N_\theta\right)$. The fact that the algorithm complexity is higher in $N_r$ than in $N_\theta$ comes from the fact that only fast Fourier transforms are performed along $\theta$ while the modal transform needs to deal with matrix vector multiplications in the radial domain. If one wants to evaluate how the code behaves with the number of modes involved $N_{\textrm{mode}}$, this can be done by setting $N_\theta=N_r$, which leads to a complexity
$\mathcal{O}\left(N^{3/2}_{\textrm{mode}}\right)$, which is far more efficient than algorithms solving the MM-GNLSE, known to behave as $\mathcal{O}\left(N^4_{\textrm{mode}}\right)$ \cite{WrightSimul,Poletti2008}.
\section{Numerical example: angular modulation instability and angular breather-like generation in ring-core fiber}
\label{subsec:example}
Nonlinear effects in ring-core fibers have become a recent subject of active study, leading to effects never observed in conventional fibers \cite{Liu,Dacha}. Here, using the code presented in this paper, we predict that, first, vector modulation instability takes place in ring-core-fibers, and second, that angular breather-like solutions can emerge in appropriate conditions. For this purpose, we consider the propagation of a continuous ($\lambda_0=1.035\mu$m) laser field propagating in the ring-core fiber considered in section \ref{RingCoreMode}. The effective mode area for the transverse electric mode is approximately 150\,$\mu$m$^2$ and the associated effective nonlinear coefficient is $\gamma\simeq 1.3$\,W$^{-1}$km$^{-1}$. The initial field $\overrightarrow{\xi}(r,\theta,z=0)$ is coupled in the transverse electric (fundamental) mode ($J=0$) with a power $P_0$=200$\,$kW.
\begin{equation}
\overrightarrow{\xi}(r,\theta,z=0)=\sqrt{I_0}F(r)\overrightarrow{e}_\theta,
\end{equation}
where $F(r)$ is the normalized radial profile of the transverse electric mode of the ring-core fiber. The peak intensity $I_0$ is defined so that
\begin{equation}
\iint{r\overrightarrow{\xi}\cdot\overrightarrow{\xi}^*drd\theta}=P_0.
\end{equation}
and approximately equals to $185\,$GW/cm$^2$. A weak random noise (both in phase and amplitude) is superimposed in the modal space to the initial field. Note that such an initial condition is experimentally realistic with femtosecond, picosecond and even nanosecond Ytterbium doped fiber amplifiers \cite{Torruellas} now commercially available. In particular, the use of long pulses will limit effects, such as self-phase modulation and modal group velocity dispersion, that could be potentially detrimental for observing the effect presented below. In order to limit the computational time, we discarded the time dependence of the field.
\begin{figure}
    \begin{center}
\includegraphics[width=9cm,keepaspectratio]{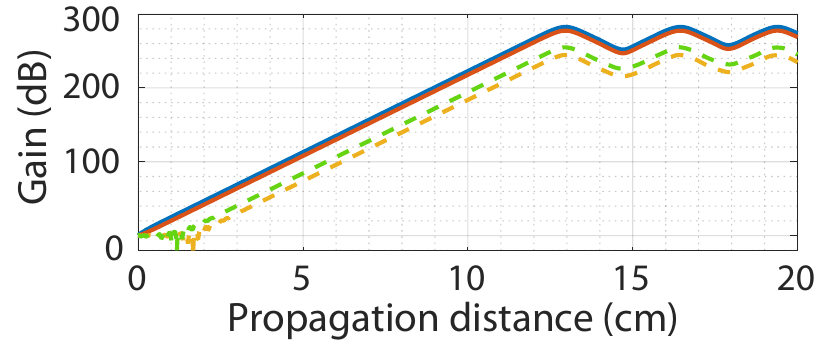}
	\caption{Energy gain of the modes $j_\ell=\pm1$ as a function of the propagation distance. The blue (resp. red) solid line corresponds to the mode $j_\ell=1$ (resp. $j_\ell=-1$) of the first group, i.e., belonging to the vectorial group of the pump. The green (resp. yellow) dashed line depicts the mode $j_\ell=1$ (resp. $j_\ell=-1$) belonging to the second vectorial group of modes.}
 	\label{Fig5}
 \end{center}
\end{figure}
\begin{figure*}[t!]
    \begin{center}
\includegraphics[width=18cm,keepaspectratio]{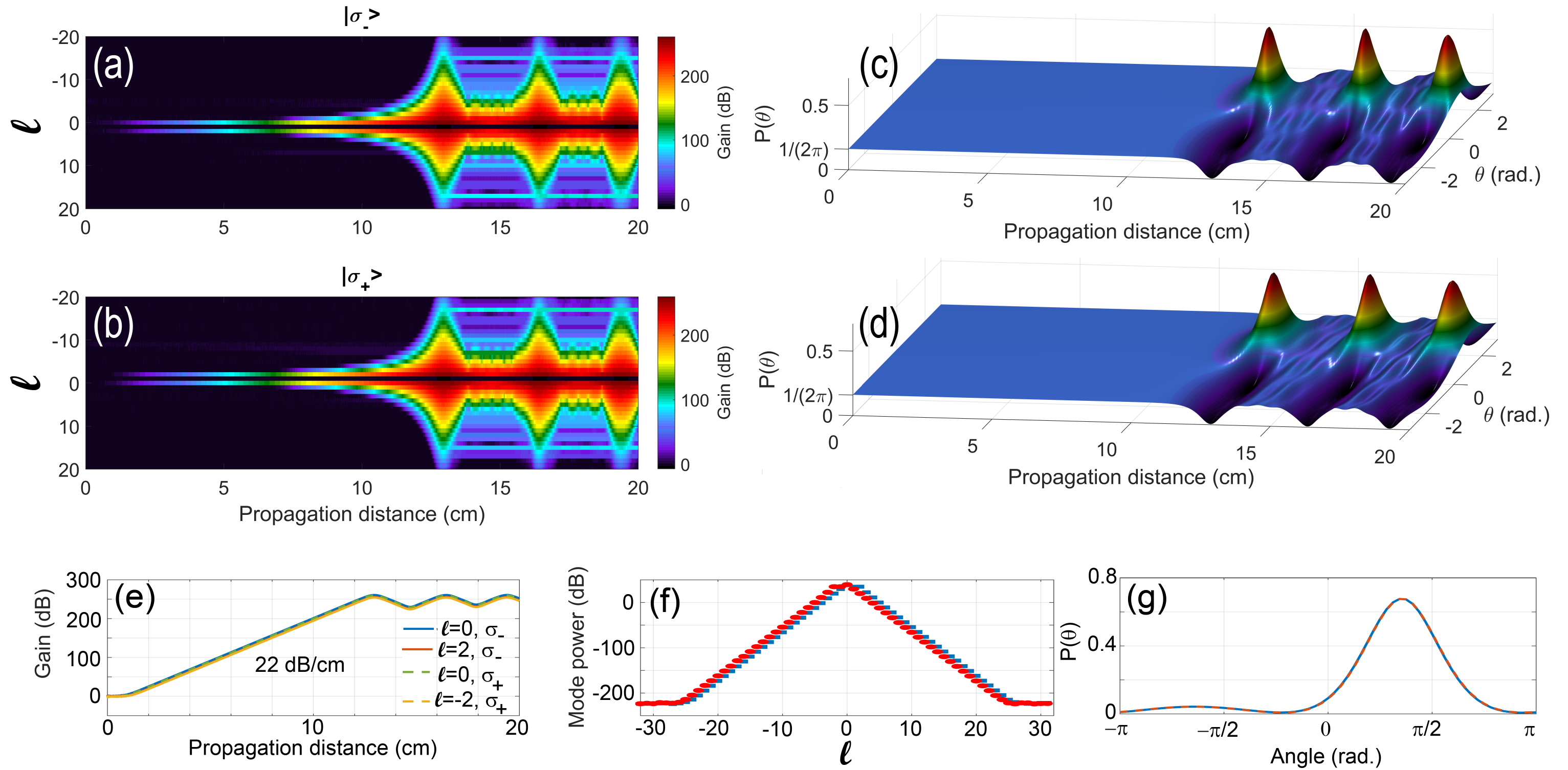}
	\caption{Orbital angular momentum spectrum (a) [resp. (b)] of the left (resp. right) circularly polarized component of the field as a function of the propagation distance. Angular distribution of the intensity profile (c) [resp. (d)] of the left (resp. right) circularly polarized component of the field as a function of the propagation distance. Energy gain of the fourth components growing during the first stage of propagation (e). Angular spectrum (f) of the left (blue squares) and right (red circles) circular components at the distance where the angular compression is maximal ($z\simeq$13\,cm) and associated angular distribution (g).}
 	\label{Fig6}
 \end{center}
\end{figure*}
At the early stage of propagation, only two particular modes start to exponentially grow up. These modes belong to the pump beam vectorial group and have a TAM $j_\ell=\pm1$ (see Fig.\,\ref{Fig5}). This indicates that only TAM-conserved processes of kind $2J_{\textrm{pump}}\rightarrow J_1+J_2$ take place at the very first stage of the propagation. Quickly after ($\simeq$1\,cm), the two modes with same total angular momenta ($j_\ell=\pm1$) but belonging to the second vectorial group starts to grow with more or less the same gain ($\simeq$22\,dB/m) than the one experienced by the two first modes. During the 7 first centimeters of propagation, only these four modes are populated. If one now decomposes the field in left and right circularly polarized components and looking at their orbital angular momentum spectra [see Figs.\,\ref{Fig6}(a,b)], only four contributions experiences gain ($\ket{\ell=0,\sigma_-}$, $\ket{\ell=2,\sigma_-}$,
$\ket{\ell=0,\sigma_-+}$, and $\ket{\ell=-2,\sigma_+}$) during the first 7\,cm. During this first step, the angularly-resolved power distribution remains flat [see Figs\,\ref{Fig6}(c,d)]. Even if a rigorous theoretical study of this process is beyond the scope of the present paper, all these observations nevertheless well correspond to an angular analogue of the (time) modulation instability process taking place in single mode fibers in the anomalous dispersion regime \cite{BejotMI}. The analogy is corroborated by noticing that the second (discrete) derivative of the propagation constant with respect to the TAM of the modes [Fig.\,\ref{Fig1}(c)] is negative. In this regard, it is particularly interesting to note that the numerical gain experienced by the two first modes very well matches the analytical formula
\begin{equation}
G(j_{\ell})=\frac{20}{\textrm{log}10}\sqrt{-\frac{k_{j_\ell}^{(2)}}{2}j_{\ell}^2\left(\frac{k_{j_\ell}^{(2)}}{2}j_{\ell}^2+2\gamma P_0\right)},
\label{EqGain}
\end{equation}
where $k_{j_\ell}^{(2)}$ is the second-order (discrete) derivative of the propagation constant with respect to $j_\ell$ [see Fig.\,\ref{Fig1}(c)] evaluated for the transverse electric fundamental mode and where $G$ is the energy gain evaluated in dB/m. Equation\,\ref{EqGain} is the angular analogue of the gain formula for modulation instability in the frequency domain. However, note that this formula is established only empirically. Further studies will be obviously needed so as to confirm or infirm its validity. The fact that angular modulation instability seems to take place suggests that angular analogue of solitons and breathers \cite{KiblerBreather,DudleyBreather} could be also observed. Note that such a possibility is made possible because there is only one single radial guided mode in such kind of fibers, which extremely simplifies the physical process. In this regard, the propagation of the initial field has been continued up to 20\,cm. From approximately 10\,cm, the orbital angular momentum spectrum of the field starts to strongly broaden for becoming maximal at a distance $\simeq 13$\,cm. At this particular distance, the gain saturates and the OAM spectrum of both circularly polarized components exhibits a triangular shape in logarithmic scale [Fig.\,\ref{Fig6}(f)]. In the direct space, this \textit{spectral} broadening is associated to an extreme angular localization of the power [Fig.\,\ref{Fig6}(g)], whose shape turns out to be very similar to breather structures already observed in fibers in the time domain. Afterwards, the angular dynamics of both polarization components exhibits multiple periodic breathing. Here again, this latter dynamics is extremely similar to those already studied in time \cite{Erkintalo}. However, it is interesting to note that the periodic and finite nature of the angular dimension over which the dynamics takes place nevertheless implies some differences in behavior with their time-domain counterparts. For instance, the fact that the orbital angular momentum spectrum is discrete by nature implies that the modulation instability process has a power threshold below which the effect vanishes. This discrete nature also explained why angular breathers appear in the above example without the use of a deterministic and coherent seeding as needed for their time-domain analogues.
\section{Conclusion}
To conclude, an efficient numerical algorithm for solving the vectorial nonlinear unidirectional pulse propagation equation has been presented. The algorithm, valid beyond the weak guidance approximation for circularly symmetric waveguides, takes advantage of the TAM conservation for being able to calculate the vectorial optical modes. The modal calculation  is used in turn so as to build a fast modal transform, which allows to easily switch from a representation of the field in the direct space to its decomposition in the modal space (and \textit{vice versa}). The presented algorithm has a complexity that roughly behaves as $N_{\textrm{mode}}^{3/2}$, which is far more efficient than conventional algorithms that scale as $N_{\textrm{mode}}^{4}$. As a first numerical example, we considered the nonlinear propagation of a moderately intense transverse electric field within a ring-core fiber. It was first shown that angular modulation instability is expected to take place, suggesting in turn the possibility to observe angular analogue of temporal solitons and breathers. This expectation has been then corroborated by exhibiting a realistic case where vectorial breather-like structures seem to emerge. While a rigorous study of such a phenomenon is out of the scope of the present paper, the perspective to observe angular breathers in ring-core fibers could nevertheless open the door to the study of new kind of nonlinear solutions. Beyond the presented monochromatic case, an extremely rich family of spatio-temporal (time+angle) dynamics could be also studied in such kind of systems. In particular, it could be interesting to see if bi-dimensional breather solutions can be found, either in the normal or anomalous dispersion (frequency) regimes. Note that since the number of angular modes is finite, we anticipate that the quest to find analytical bi-dimensional breather solutions in such a case will be far more tractable than in the general 3D+1 situation where both radial and angular dimensions have to be taken into account. Note, however, that the fact that optical modes are strongly vectorial could nevertheless complicate a bit the theoretical study.
\section*{Acknowledgments}
The author acknowledges financial support of French programs "Investments for the Future" operated by the National Research Agency (ISITE-BFC, contract ANR-15-IDEX-03; EIPHI Graduate School, contract ANR-17-EURE-0002; EQUIPEX+ Smartlight, contract ANR-21-ESRE-0040), from Bourgogne Franche-Comt\'e region and European Regional Development Fund. The author thanks the CRI-CCUB for CPU loan on the multiprocessor server.
\section*{Disclosures}
The author declares no conflicts of interest.
\section*{Data availability}
Data underlying the results presented in this paper are not publicly available at this time but may be obtained from the authors upon reasonable request.

\end{document}